\documentstyle[aps,prd,mathrsfs]{revtex}
\def\Tr{\mbox{Tr}\,}
\def\W{\wedge}
\def\dag{^\dagger}
\def\D{\mbox{d}}

\def\half{\frac{1}{2}}

\def\slash#1{\, /\kern-0.6em{#1}}

\begin{document}
\tighten

\preprint{{hep-th/0107104}}

\twocolumn[\hsize\textwidth\columnwidth\hsize\csname
  @twocolumnfalse\endcsname

\title{Gauge transformations of the non-Abelian two-form}     
\author{Amitabha Lahiri}
\address{S. N. Bose National Centre for Basic Sciences, \\
Block JD, Sector III, Salt Lake, Calcutta 700 098, INDIA}
\address{amitabha@boson.bose.res.in}
\maketitle

\begin{abstract}
A novel inhomogeneous gauge transformation law is proposed for a
non-Abelian adjoint two-form in four dimensions. Rules for
constructing actions invariant under this are given. The auxiliary
vector field which appears in some of these models transforms like
a second connection in the theory. Another local symmetry leaves
the compensated three-form field strength invariant, but does not
seem to be generated by any combination of local constraints.  Both
types of symmetries change the action by total divergences,
suggesting that boundary degrees of freedom have to be taken into
account for local quantization.
\end{abstract}

\pacs{PACS: \,03.50.Kk, 11.30.Ly, 12.60.Cn \\
Keywords: Topological field theory, non-Abelian two-form,
non-local symmetry\, }

\medskip
  ]


The non-Abelian two-form was first introduced in the literature in
the context of nonlinear sigma models~\cite{Freedman:1977pa} via an
interaction term of the form $\Tr \epsilon^{\mu\nu\rho\lambda}
B_{\mu\nu}F_{\rho\lambda}$, with $B$ a two-form potential living in
the adjoint representation of the gauge group, and $F$ the field
strength of the gauge field. An action made up of this term alone
is a Schwarz-type topological field theory~\cite{Schwarz:1978cn}, a
four dimensional generalization of the Chern-Simons action, with no
perturbative degree of freedom.  Since this action does not depend
on the metric of the background space-time, it has been suggested
that it could be thought of as a toy model for some features of
quantum gravity~\cite{Horowitz:1989ng}.

It was known from the beginning~\cite{Freedman:1977pa} that this
interaction could lead to a massive vector field in four dimensions
if a suitable quadratic (kinetic) term for $B$ could be found. This
would generalize the corresponding mechanism for Abelian vector
fields~\cite{Cremmer:1974mg}, which does not have the Higgs
particle in the spectrum. An action which maintains the symmetries
of the interaction term, and reduces to the Abelian mechanism, was
found sometime ago \cite{Lahiri:1992hz}.  Although it requires a
non-propagating auxiliary vector field which decouples in the
Abelian limit, some algebraic methods of quantization of gauge
theories seem to be applicable~\cite{Hwang:1997er,Lahiri:2001uc},
so it may be useful as a Higgs-free mass mechanism for vector
fields. Another quadratic term involving the same auxiliary vector
field was used recently for a first order formulation of Yang-Mills
theory~\cite{Cattaneo:1998eh}. The non-Abelian two-form also makes
its appearance in a loop space formulation of Yang-Mills
theory~\cite{Chan:1995bp}, as well as in the loop space formulation
of gravity as a gauge theory~\cite{Smolin:2000qp}.

Despite the wide applicability of two-form field theories, the
nature of the field remains obscure, or at best poorly
understood. Geometrically it is natural to think of $B$ as a gauge
field for (open or closed) strings, since a two-form couples to a
surface. This is a consistent picture for the Abelian two-form, but
not for its non-Abelian counterpart, because of the difficulty in
defining `surface ordered' exponentials, which appear when a
two-form is coupled to the world surface of a string, in a
reparametrization invariant fashion~\cite{Teitelboim:1986ya}.  The
algebraic counterpart of this result is that there is no simple
non-Abelian generalization of the symmetries and constraints of
dynamical two-form theories~\cite{Henneaux:1997mf}. In this Letter
I display some novel {\em classical} local symmetries of these
theories. Some of these symmetries, although dependent on arbitrary
local parameters, cannot be implemented by local constraints.
Existence of such symmetries needs a fundamentally new paradigm for
quantization of field theories in general, and thus requires a
review of established results for the two-form in particular.

It will be convenient to use the notation of differential forms.
The gauge connection one-form (gauge field) is defined in terms of
its components as $A = -ig A^a_\mu t^a \D x^\mu\,,$ where $t^a$ are
the generators of the gauge group satisfying $[t^a, t^b] =
if^{abc}t^c$ and $g$ is the gauge coupling constant. Any other
coupling constant, which may be required in a given model, will be
assumed to have been absorbed in the corresponding field. This will
cause no problem since I am dealing with classical systems.  The
gauge group will be taken to be SU(N). The gauge-covariant exterior
derivative of an adjoint $p$-form $\xi_p$ will be written as
\begin{eqnarray}
\D_A\,\xi_p \equiv \D\,\xi_p + A\W \xi_p + (-1)^{p+1} \xi_p\W A\,,
\end{eqnarray}
where $\D$ stands for the usual exterior derivative. The field
strength is $F = \D A + A\W A$, and satisfies Bianchi identity, $\D
F + A\W F - F\W A = 0$.  Under a gauge transformation, the gauge
field transforms as $A \to A' = UAU\dag +\, \phi$, where for later
convenience I have defined $\phi \equiv - \D UU\dag$. Note that
$\phi$ is a flat connection, $\D\phi + \phi\W\phi = 0$.
 
All known theories of the non-Abelian two-form are built around the
`topological' coupling term {$\int\Tr B\W F$}, so any discussion of
symmetries must start with the invariance of this term, which will
be called the action in what follows. Quite obviously the
action is invariant under gauge transformations if $B$ transforms
homogeneously in the adjoint, $B\to B' = UBU\dag$. In addition, the
action is invariant under a non-Abelian generalization of
Kalb-Ramond gauge transformation~\cite{Kalb:1974yc},
\begin{eqnarray}
B' = B + \D_A\xi, \qquad A' = A\,,
\label{0501.vector}
\end{eqnarray}
where $\xi$ is a one-form which vanishes sufficiently rapidly at
infinity. This will be referred to as vector gauge transformation,
while a transformation with $U$ will be called (SU(N), or usual)
gauge transformation. The two transformations combine as
\begin{eqnarray}
A' = UAU\dag + \phi\,,\qquad B' = UBU\dag + \D_{A'}\xi'\,,
\label{0501.combo1}
\end{eqnarray}
if $\xi$ transforms homogeneously in the adjoint, $\xi'
= U\xi U\dag$.

All this is well known, any BRST analysis of the non-Abelian
two-form uses these transformation rules. In this Letter I will
display a new set of transformations. First of all, note that the
action is invariant under Eq.~(\ref{0501.combo1}) even if $\xi$
transforms like a connection, but then vector gauge transformations
do not form a group, since $\xi_1 + \xi_2$ is not a connection if
$\xi_1$ and $\xi_2$ are. Otherwise $\xi$ is (almost)
arbitrary. This arbitrariness can be traded for choosing a
connection in place of $\xi'$ in Eq.~(\ref{0501.combo1}), by
specifically choosing the flat connection $\phi = -\D UU\dag$,
\begin{eqnarray}
A' &=& UAU\dag + \phi\,,\nonumber \\
B' &=& UBU\dag +  UAU\dag\W\phi + \phi\W UAU\dag +
\phi\W\phi\,.  
\label{0501.combo}
\end{eqnarray}
Since $\phi$ is completely determined by the SU(N) gauge
transformation $U$, this equation has absolutely nothing to do with
the original vector gauge transformation --- no arbitrary vector
parameter appears in the transformation rule. Instead I can now
think of this equation as a modification of the SU(N) gauge
transformation rule of $B$. This statement should be qualified,
since there is a further complication --- Eq.~(\ref{0501.vector})
left the action invariant because $\xi$ vanished at the boundary,
while the gauge transformation matrix $U$, and therefore the flat
connection $\phi$, need satisfy no such condition. Consequently the
action is invariant only up to a total divergence,
\begin{eqnarray}
\int \Tr B'\W F'
&=& \int \Tr  B\W F + \D\,\Tr\,(\phi\W UFU\dag) \,,
\label{0501.deltaS}
\end{eqnarray}
where I have used cyclicity of trace, Bianchi identity, and the
fact that $\phi$ is flat. Variation by a total divergence under
gauge transformations reinforces the picture of the four
dimensional $B\W F$ model being a generalization of three
dimensional Chern-Simons theory. Note that Eq.~(\ref{0501.combo})
is trivial in the Abelian limit, and is a symmetry of the action
only if $B$ is a two-form and $A$ is a one-form, and therefore only
in four dimensions.

The appearance of the gauge connection $A$ in the gauge
transformation rule for $B$ may seem a little odd at first sight,
but remember that these are objects in a theory with local SU(N)
symmetry.  If the objects are dynamical, the theory must contain
derivatives, and therefore a connection. This need not be the
dynamical gauge field $A$, but may be a flat connection (gauge
transform of the situation with no $A$), as is used in the naive
generalization of the duality relation between a two-form and a
scalar~\cite{Oda:1990tp}, or outside the horizon of black holes
with a non-Abelian topological charge~\cite{Lahiri:1992yz}. In such
cases this symmetry is exact, since $F=0$. ($B\W F$ also vanishes
in these cases, but there are other terms in the theory.)

Away from flat connections, the divergence term gives a surface
integral which vanishes if either $F \to 0$ or $\phi \to 0$
sufficiently rapidly on the boundary. The first can happen if for
example $F$ represents a finite action configuration of Yang-Mills
theory, while the second can happen only if $U \to 1$ on the
boundary.  When the surface integral does not vanish, the classical
conserved current of local SU(N) symmetry for the pure $B\W F$
theory has a {\em topologically conserved} component proportional
to ${}^*\D F\sim\epsilon^{\mu\nu\rho\lambda} \partial_\nu
F^a_{\rho\lambda}$. This component of the conserved current is not
gauge covariant in general, but in a configuration where $F$
vanishes on the boundary (e.g. Euclidean finite action) this makes
a vanishing contribution to the conserved charge.

It is necessary to check that the `new and improved' SU(N) gauge
transformation rule for $B$ obeys the group composition law. 
Indeed, if $U_1$ and $U_2$ are two SU(N) transformations applied
successively, let 
\begin{eqnarray}
B_1 &=& U_1BU_1\dag + U_1AU_1\dag\!\W\phi_1 + \phi_1\!\W
U_1AU_1\dag + \phi_1\!\W\phi_1\,
\end{eqnarray}
in obvious notation. Then Eq.~(\ref{0501.combo}) is recovered with
$U = U_2U_1$ and $\phi = -\D UU\dag.$

The BRST transformations corresponding to Eq.~(\ref{0501.combo}),
with indices and coupling constants restored, are
\begin{eqnarray}
s A^a_\mu &=&  \partial_\mu\omega^a +
gf^{abc}A^b_\mu\omega^c\,,\nonumber \\
s B^a_{\mu\nu} &=& gf^{abc} B^b_{\mu\nu}\omega^c +
gf^{abc}A^b_{[\mu}\partial_{\nu]}\omega^c\,, \nonumber\\
s \omega^a &=& -\half gf^{abc}\omega^b\omega^c\,.
\label{0501.BRST1}
\end{eqnarray}
This is nilpotent, $s^2 = 0$. Comparison with the conventional BRST
rules
\cite{Hwang:1997er,Lahiri:2001uc,Cattaneo:1998eh,Thierry-Mieg:1983un}
shows that the second term in $s B^a_{\mu\nu}$ is like the vector
gauge transformation with $\partial_\mu\omega^a$ playing the role
of the vector parameter, but this is not the whole story, nor a
true one, as I will discuss below. Now I make a small diversion.

Even without vector gauge symmetry, various actions are invariant
under the new and improved gauge transformation of the two-form.
Note that
\begin{eqnarray}
\D A' = -\phi\W UAU\dag + U\D AU\dag - UAU\dag\W\phi + \D\phi\,,
\end{eqnarray}
so that $B + \D A$ transforms covariantly under
Eq.~(\ref{0501.combo}), $B' + \D A' = U(B + \D A)\,U\dag$. So an
interaction term of the form $\Tr(B + \D A)\W F$ will be exactly
invariant (and the conserved current for SU(N) gauge
transformations will be gauge covariant). The extra term in the
interaction is a total divergence, $\Tr\,\D A\W F = \D\,\Tr (A\W\D
A + \frac13 A\W A\W A)$. This is {\em not} the well known
Chern-Simons term, the numerical factor on the second term is
different.  (For a Euclidean finite action configuration the
integral of this term is still proportional to the Pontryagin
index.)  I can now construct several novel actions involving only
$A$ and $B$, with interesting physical implications, by demanding
invariance only under the gauge transformation of
Eq.~(\ref{0501.combo}). The starting point is either $\Tr B\W F$,
which is invariant up to a total divergence, or $\Tr(B + \D A)\W
F$, which is exactly invariant. To either of this a quadratic term,
constructed out of $(B + \D A)$, or the covariant field strength
$(\D_A B - \D F)$, can be added for an invariant action. Finally,
note that while Teitelboim's proof~\cite{Teitelboim:1986ya} has
{\em not} been invalidated, the results here suggest that the
object which couples to a string is $(B + \D A)$ or something
similar. The integral of this over the world surface will have a
contribution from line integrals of $A$ over initial and final
configurations, and for open strings over the world lines of end
points. So it may be worthwhile to reinvestigate that proof for a
better understanding of the geometrical nature of two-forms.
Similar comments apply to other results about the quantum
non-Abelian two-form. End of diversion.

None of the actions mentioned in the above paragraph is invariant
under vector gauge transformations.  These need to be addressed
separately.  These transformations leave invariant the interaction
term $\int\Tr B\W F$ (if the gauge parameter $\xi$ vanishes
sufficiently rapidly on the boundary), but cause problems for any
term quadratic in $B$. Indeed a theorem~\cite{Henneaux:1997mf}
asserts that a kinetic term for the two-form, invariant under both
types of gauge transformations, cannot be constructed unless
additional fields are introduced to compensate for the vector gauge
transformations.

This was the route taken earlier in the non-Abelian generalization
of the mass generation mechanism for vector
fields~\cite{Lahiri:1992hz}, where an auxiliary one-form $C$ was
introduced, and a compensated field strength $H$ was constructed as
$H = \D_A B - F\W C + C\W F$.  With the vector gauge transformation
law modified to include $C$,
\begin{eqnarray}
A' = A\,,\qquad B' = B + \D_A\xi\,,\qquad C' = C + \xi\,,
\label{0501.vector2}
\end{eqnarray}
this field strength remained invariant and its square could provide
the kinetic term for $B$. This one-form was also introduced in the
`improved' first-order formulation of Yang-Mills
theory~\cite{Cattaneo:1998eh}, where instead of a kinetic term, a
quadratic invariant $\Tr(B + \D_A C)^2$ was added to the action.
Just as for the mass generation mechanism, the auxiliary field
helps retain the vector gauge symmetry.

Now suppose the two-form is taken to transform under the
non-Abelian gauge group as in Eq.~(\ref{0501.combo}), and symmetry
under vector gauge transformations is required. Then the auxiliary
field $C$ must transform like a connection under usual gauge
transformations,
\begin{eqnarray}
C' = UCU\dag + \phi\,,\qquad s C^a_\mu = \partial_\mu\omega^a +
gf^{abc}C^b_\mu\omega^c\,,
\label{0501.ctrans}
\end{eqnarray}
and the combination $B - \D_A C$ can be used for construction of
invariant actions, just as it has been
in~\cite{Lahiri:1992hz,Hwang:1997er,Lahiri:2001uc,Cattaneo:1998eh},
but now with the new rules for gauge transformations. This is the
second result of this Letter. For vector gauge transformations,
Eq.~(\ref{0501.vector2}) holds, with $\xi$ transforming
homogeneously in the adjoint. The comment made after
Eq.~(\ref{0501.BRST1}) should be obvious now. Since $\phi$ does not
transform homogeneously under the gauge group, the new rules for
gauge transformations cannot be a special case of vector gauge
transformations, despite the formal similarity. Nor is it possible
to take the vector parameter $\xi$ to transform like a connection,
because connections do not add, so the gauge transformation
properties of $C$ will be ill-defined. This also provides an
additional reason why $C=0$ is not a `good gauge choice' : $(C +
\xi)$ transforms like a connection, so even if it is made to vanish
in one gauge, it will be non-zero in another.

This comment leads to a local symmetry of the three-form field
strength $H$, and therefore of the massive vector model, with
far-reaching consequences. The vector gauge parameter $\xi$ has to
transform homogeneously in the adjoint, i.e., like the difference
of two connections.  So let me choose an arbitrary SU(N) matrix
valued field $\tilde U(x)$ (not the gauge transformation of
Eq.~(\ref{0501.combo})) and write $\tilde\phi = -\D\tilde U\tilde
U\dag$.  Using this, I postulate the transformations
\begin{eqnarray}
C' &=& C +\alpha (A - \tilde\phi)\,, \quad A' = A \,,\nonumber \\
B' &=& B +\alpha (A - \tilde\phi)\W(A - \tilde\phi) \,,
\label{0501.special}
\end{eqnarray}
with $\alpha$ an arbitrary constant.  The $B\W F$ term changes by a
total divergence, and the compensated field strength $H$ is
invariant, while $B - \D_A C$ is not, i.e.,
\begin{eqnarray}
\delta\int\Tr B\W F &=& \alpha \int\D\,\Tr (\frac13 A\W A\W A -
\tilde\phi\W F)\,,\nonumber \\
\delta(B - \D_A C) &=& - \alpha F\,,\qquad \delta H = 0\,.
\label{0501.nonpert}
\end{eqnarray}
Note that these are {\em not} a special case of the vector gauge
transformations, even if it appears as if they were. Firstly, the
local transformation (scalar) parameters, present in $\tilde\phi$,
need not vanish at infinity. This is actually a red herring, since
I can always choose $\tilde U$ so that $\tilde\phi$ vanishes at
infinity. But the gauge field $A$ need not vanish at infinity, so
this is distinct from the usual vector gauge transformations. 

There is a more important difference. I could think of
Eq.~(\ref{0501.special}) as local transformations, since $\tilde U$
is an arbitrary local SU(N) matrix. In general local
transformations are generated by local constraints. The constraints
of theories with a dynamical two-form are known~\cite{Lee:1998qu},
including those which generate vector gauge transformations, and it
is immediately obvious that they do not generate
Eq.~(\ref{0501.special}). Further, it does not seem possible to
incorporate these transformations into the BRST generator, unlike a
transformation of the type $\delta B = - \alpha
F$~\cite{Lahiri:1992yz,Lahiri:1992hz}. This last can be used for
eliminating unwanted terms, and may be incorporated in the BRST
construction by using a constant anticommuting
ghost~\cite{Lahiri:2001uc}. A similar transformation for the
Abelian two-form was investigated in~\cite{Deguchi:1999xp}, but
because the field strength there included the Chern-Simons
three-form as in string theory, the corresponding ghost was not a
constant. These approaches will not work for the transformations of
Eq.~(\ref{0501.special}). These should properly be called {\em
semiglobal} transformations --- a combination of global and local
transformations --- which depend on a global parameter $\alpha$ and
local parameters in $\tilde U$. The {\em global} aspect should be
emphasized because these are connected to the identity
transformation in the limit $\alpha \to 0$ but not in the limit
$\tilde U \to 1$. It seems unlikely that standard methods of
perturbative quantization of gauge theories will be able to protect
symmetries like this. This is the third and final result of this
Letter.

Let me make a final observation about Eq.~(\ref{0501.special}). To
what extent does the change in the action depend on the boundary
values of the fields? Suppose I consider the system in a background
where $A$ goes to a pure gauge at infinity, and choose $\tilde U$
such that $\tilde\phi \to A$ at infinity. Then the boundary values
of all fields $A, B, C,$ are unaffected by these transformations.
Yet the local transformation still modifies the action by a
boundary integral. In other words, just as the local symmetries of
the system is dependent on the boundary behavior of the fields,
that behavior is affected by local transformations.

The surfeit of actions and transformations mentioned in this Letter
should not detract attention from its central result, which is that
Eq.~(\ref{0501.combo}) is a local, {\em classical}, SU(N) gauge
symmetry of the four dimensional non-Abelian $B\W F$ theory. This
symmetry exists only in four dimensions, there is no analog in
lower or higher dimensions, or for Abelian theories. For quantum
theories built around this term, it seems possible to include this
symmetry in a perturbative analysis (see Eqs.~(\ref{0501.BRST1})
and (\ref{0501.ctrans})). A secondary result appears when an
auxiliary connection is introduced, as is done in various
models. The compensated field strength is invariant under a novel
local classical transformation, shown in Eq.~(\ref{0501.special}),
which cannot be implemented by any combination of constraints. It
has been suggested~\cite{Amorim:1999mr,Lahiri:2001uc} that
canonical methods of quantization are inferior to functional
methods for theories of the dynamical two-form, and may even be
inapplicable. While it is not clear if functional methods of
quantization will preserve these symmetries in some form, it seems
very unlikely that canonical methods will be able to do so at
all. I end this Letter with the suggestion that theories of the
dynamical non-Abelian two-form provide a ready testing ground for
different methods of field quantization in four dimensions,
especially where boundary degrees of freedom and topological
`quantum numbers' are involved.







\end{document}